**Directional commensurability stabilizes structural superlubricity in patterned mesoscale interfaces**

*Viet Hung Ho[#], Ge Li[#,*], Melisa M. Gianetti, Bjørn Haugen, Graham L. W. Cross, Astrid S. de Wijn[*]*

Viet Hung Ho, Ge Li, Melisa M. Gianetti, Bjørn Haugen, Astrid S. de Wijn

1. Department of Mechanical and Industrial Engineering, Norwegian University of Science and Technology (NTNU), 7491, Trondheim, Norway

Graham L. W. Cross

2. School of Physics and CRANN, Trinity College Dublin, Dublin 2, Ireland

# These authors contributed equally

* Corresponding Author

Email: astrid.dewijn@ntnu.no, ge.li@ntnu.no

**Abstract:**

Structural superlubricity, arising from lattice incommensurability, offers a promising route to eliminate friction and associated energy losses in mechanical systems. In real-world systems, roughness and wear currently pose severe limitations on its robustness and especially the contact size. Here, we consider patterned surfaces as a possible route to overcome some of these limitations. We show that the simplest choice of patterning, contacts made up of two incommensurate triangular-triangular patterns, fails at elevated loads because of the small number of load-bearing contacts, causing the maximum local contact pressure to exceed the strength of the superlubric coating. We introduce a square-triangular patterned interface that increases the number of load-bearing contacts and organizes them into continuous contact lines. When sliding along specific directions relative to these lines, superlubricity is maintained at significantly higher loads by reducing pressure-induced coating failure while also remaining somewhat tolerant to surface imperfections. These findings establish a mechanism for stabilizing structural superlubricity against coating failure and a design principle for engineering low-friction interfaces with enhanced load-bearing capacity and defect tolerance.

## 1. Introduction

Friction accounts for approximately 23% of the world's total energy consumption[1]. The development of low-friction technologies is thus critically important for improving both energy efficiency and environmental sustainability[1-2]. While there are many mechanisms that lead to friction, a large fraction of energy losses in engineering contacts originates from solid-solid contacts[3].

For this reason, achieving extremely low friction is a central objective in the field of tribology. Structural superlubricity represents one promising route, arising from the interaction between atomically flat surfaces with structurally incompatible (incommensurate) lattices. It was first predicted theoretically and subsequently observed by Hirano *et al.* through controlled tuning of the commensurability between two atomically flat mica lattices[4]. Since then, structural superlubricity has been demonstrated across a wide range of interfaces involving two-dimensional (2D) materials, including graphite[5], graphene[6], molybdenum disulfide ($MoS_2$)[7], transition-metal carbides and nitrides (MXenes)[8], diamond-like carbon (DLC)[9], and various heterostructures[10]. Despite these advances, structural superlubricity remains largely confined to carefully controlled lab conditions at the nano- and microscale[11].

In practical engineering conditions, particularly at larger contact sizes, surface roughness and wear dominate interfacial behavior[12]. As a result of surface roughness, real sliding interfaces consist not of a single atomically flat contact, but of a collection of small discrete asperities that govern the effective contact area and load distribution[13]. These concentrate the applied load into small real contact areas, producing local pressure peaks that can destabilize the interface and initiate coating failure and wear. Once this occurs, damage to the sliding interface breaks the structural superlubricity.

One route that has been suggested for dealing with the issue of multiple asperities is to use patterned surfaces with superlubric coatings[6, 14]. This patterning allows for more controlled asperity contacts and is expected to improve the tolerance for long-range roughness[15].

A particularly promising idea of patterning is to translate the ideas of structural incompatibility to the meso scale: a combination of two mesoscale patterned sliders consisting of two assemblies of asperities arranged in a well-defined lattice pattern that captures the underlying interfacial registry, in an incommensurate combination of geometries. The corrugation arising from the interaction between the two patterned surfaces is closely related to the interfacial frictional response. Smaller corrugation generally corresponds to lower friction, as the slider needs to overcome smaller barriers during sliding, if the superlubric coating remains intact. Similarly to the atomic scale, the interface between incommensurate patterns is expected to produce a smaller corrugation than one that is commensurate.

In this work, we investigate this possibility in simulations and elucidate how the geometric patterning of the contacting bodies influences contact formation, contact pressure distribution, and ultimately frictional behavior. We show that while friction can be engineered to be very low, this incommensurate double-patterning unfortunately also leads to local coating failure due to high local contact pressure, which in turn leads to a loss of superlubricity. We propose a quasi-incommensurate geometry that avoids this issue, preserving both low friction and high resistance to coating failure while remaining tolerant to realistic surface imperfections.

The interface is modeled as an assembly of coating-covered spherical particles representing microscale asperities. Specifically, we consider a substrate composed of closely packed spheres arranged in a triangular lattice, as shown in **Figure 1a**. This configuration represents the optimal packing in two dimensions and is also the structure most readily formed in experiments[16]. Sliders of two different lattice types, triangular and square, are considered and shown in Figure 1b and 1c, respectively. The superlubric coating is represented by a pressure-dependent friction model, where coating failure is triggered when the local maximum Hertzian contact pressure exceeds 4 GPa, the reported durability limit of van der Waals layered coating materials[15, 17]. While this choice is still somewhat arbitrary, it has no qualitative effect on our results. This setup enables a systematic investigation of the roles of pattern commensurability, orientational alignment, and sliding direction on friction behavior. For clarity, schematic side views of the triangular-triangular (tri-tri) and square-triangular (square-tri) models are shown in Figure 1d and 1e, respectively, and a three-dimensional representation of the spring-driven sliding system is shown in Figure 1f. Detailed parameters of the spheres forming the sliders and substrate, as well as the dimensions of both systems of two sliders, are provided in the Simulation model and methods section.

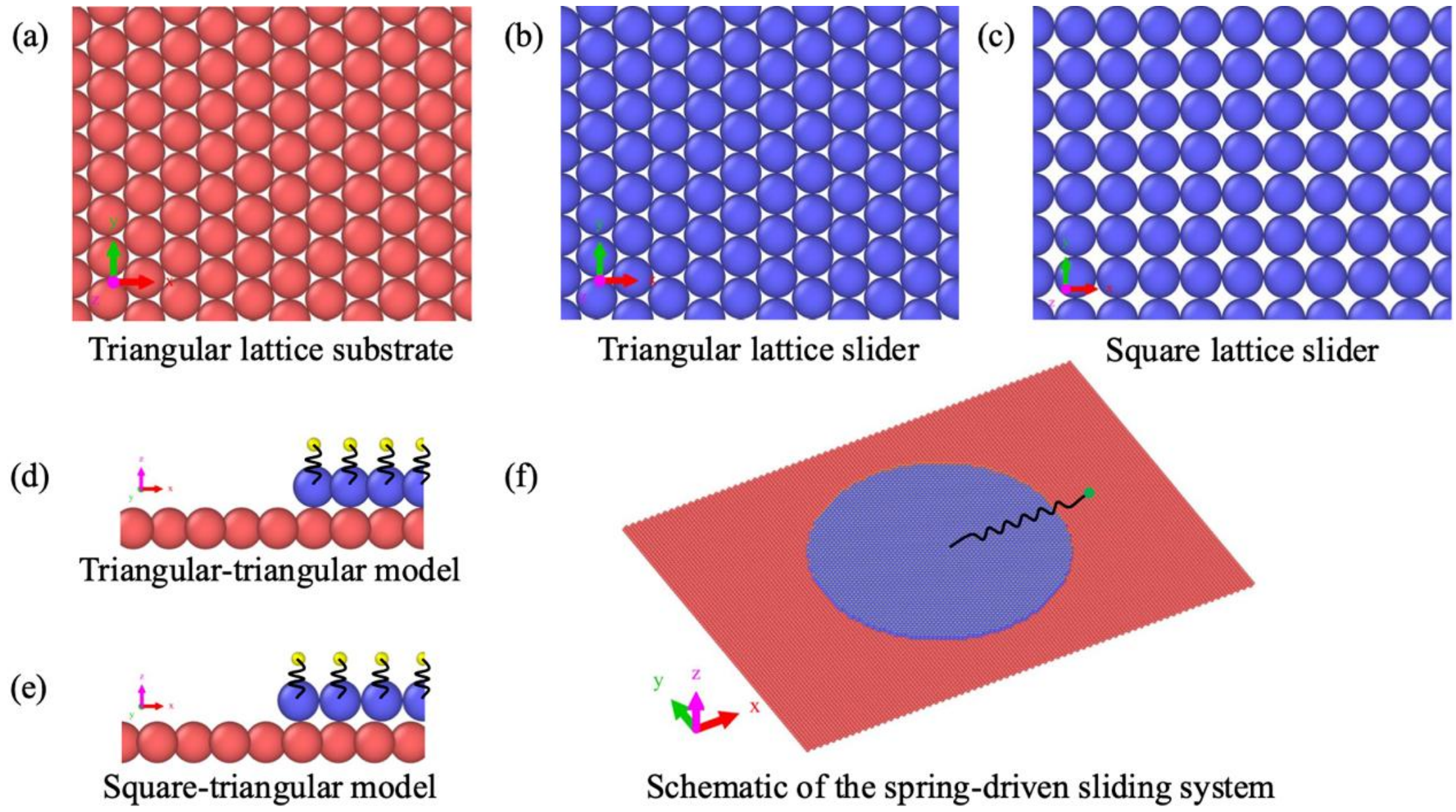

**Figure 1:** Geometric configurations of the substrate, the two slider lattices, and the corresponding model systems.

(a) Top view of a section of the substrate with red spheres arranged in a triangular lattice.

(b) Top view of a section of the triangular-lattice slider with blue spheres.

(c) Top view of a section of the square-lattice slider with blue spheres.

(d) Side view of the triangular-triangular (tri-tri) configuration, comprising a triangular-lattice substrate and a triangular-lattice slider connected to a triangular-lattice control layer through springs.

(e) Side view of the square-triangular (square-tri) configuration, comprising a triangular-lattice substrate and a square-lattice slider connected to a square-lattice control layer through springs.

(f) Schematic of the spring-driven sliding system, consisting of a circular slider in blue and a control layer in yellow on a triangular-lattice substrate in red and a support in green.

## 2. Results

Simulations were carried out for the fully incommensurate tri-tri configuration at a relative orientation angle of 30° shown in **Figure 2a** under normal loads ranging from 0.35 to 3.86 mN and the friction forces were computed. For comparison, sliding simulations were also performed of the fully commensurate tri-tri configuration at a relative orientation angle of 0°. As shown in Figure 2b, the best incommensurate configuration maintains superlubricity up to a normal load of 1.75 mN. At higher normal loads, the friction starts to rapidly increase. The configuration with fully commensurate patterns exhibits relatively high friction everywhere that increases with normal load, even under low loading conditions, as expected. Commensurate sliders at normal loads above 1.75 mN become trapped in a geometrically interlocked state between opposing asperities and no steady dynamic sliding is observed.

To understand the frictional behavior, the maximum interfacial pressure ($P_{max}$) and the total contact number between the slider and the substrate were analyzed. The results are shown in Figure 2c. For the typical incommensurate configuration at θ = 30°, the contact number increases gradually from 37 to 121 with increasing normal load, which remains insufficient to distribute the applied load evenly. Consequently, $P_{max}$ exceeds the critical strength of the superlubric coating (4 GPa) above a normal load of 1.75 mN, leading to coating failure and breakdown of the superlubric state. In contrast, the fully commensurate configuration at θ = 0° maintains large and constant 3505 contacts, allowing the applied load to be distributed more uniformly. As a result, the maximum $P_{max}$ (1.05 GPa) under the highest normal load of 3.86 mN remains well below the critical threshold, preserving the integrity of the superlubric coating. A change in the critical strength of the coating would only shift the normal force at which this failure occurs, but not lead to any qualitative differences in the behavior[15].

These results reveal the importance of the total number of contacts in the system. Although structural superlubricity can be achieved in a fully incommensurate configuration, its stability is limited under high loads because the small number of load-bearing contacts promotes high local pressure and local coating failure.

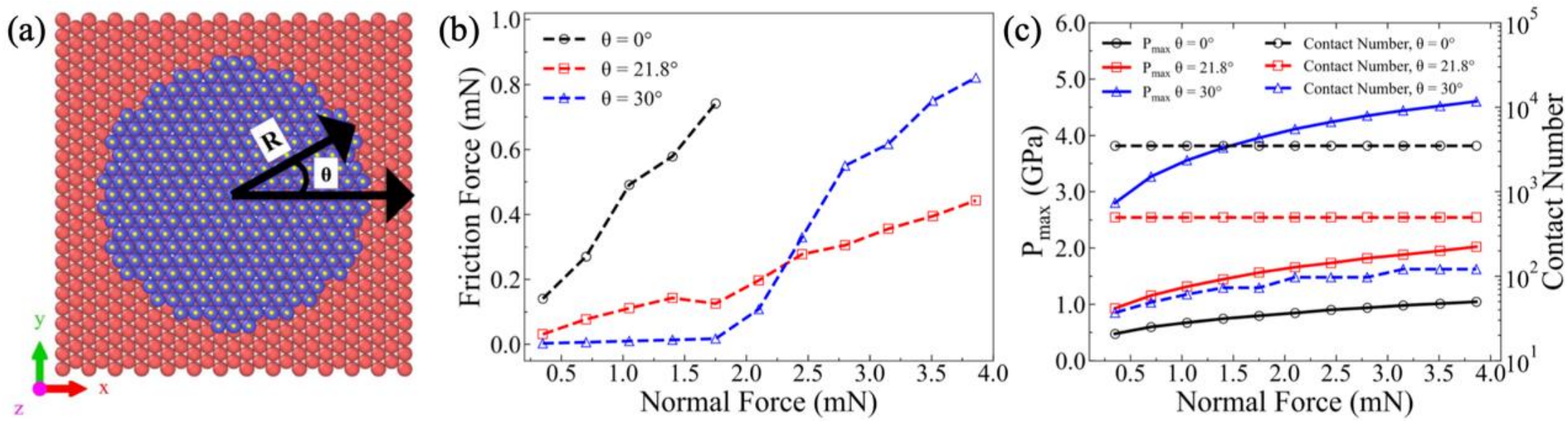

**Figure 2:** Cartoon of the triangular-triangular (tri-tri) model, frictional response, maximum contact pressure ($P_{max}$) and contact number of the tri-tri model under varying loads.

(a) Cartoon top view of the fully incommensurate tri-tri configuration. The dimensions of the surface and the orientation angle θ are indicated for the orientation study. The configuration shown corresponds to θ = 30°.

(b) Friction force, (c) $P_{max}$ and contact number of the slider for the three selected initial configurations under normal forces from 0.35 to 3.86 mN. The relative orientations of 0° and 30° correspond to the commensurate and incommensurate state, respectively. The selected orientation angle of 21.8° is discussed below.

In the tri-tri system, different contact patterns can be generated through orientational alignment of the slider. This raises the possibility of identifying configurations that balance interfacial incommensurability with a sufficiently large number of load-bearing contacts. To investigate this, a series of simulations were performed where the slider was equilibrated in the range of orientation angles of between 0 and 30° with increments of 0.1°. For each relative orientation angle, the resulting contact configuration, including the contact number and $P_{max}$, was analyzed.

**Figure 3a** and 3b present the total contact number and the maximum contact pressure ($P_{max}$) on the slider spheres for each orientational configuration under normal loads of 0.35, 1.4, 2.1, and 3.86 mN. As expected, both the contact number and $P_{max}$ increase with increasing normal load. Notably, markedly lower $P_{max}$ values are observed at several angles (13.2°, 21.8°, and 27.8°) under all applied loads, where the contact numbers are also relatively high. When the normal load is 3.86 mN, $P_{max}$ exceeds the strength of the superlubric coating (= 4 GPa) at most orientations.

As discussed above, a configuration combining a relatively large number of contacts with a certain extent of interfacial incommensurability may facilitate low friction. Therefore, we further investigate the configuration which corresponds to the maximum contact number, i.e., $\theta = 21.8°$. Together with the fully commensurate ($\theta = 0°$) and incommensurate ($\theta = 30°$) configurations discussed above, the corresponding contact configurations and pressure distributions are shown in Figure 3c, 3d and 3e. The resulting friction forces, $P_{max}$ and contact number are presented in Figure 2b and 2c. For low normal loads, the case with relative orientation 21.8° has a friction coefficient between the cases of 0 and 30°. Under higher normal loads, however, the configuration at $\theta = 21.8°$, which possesses a larger number of contacts (499), maintains relatively low friction by preserving the integrity of the superlubric coating layer. These results demonstrate that, in the tri-tri configuration, friction can be tuned through orientational alignment, although robust superlubricity can only be maintained within a limited normal-load range while the coating remains intact.

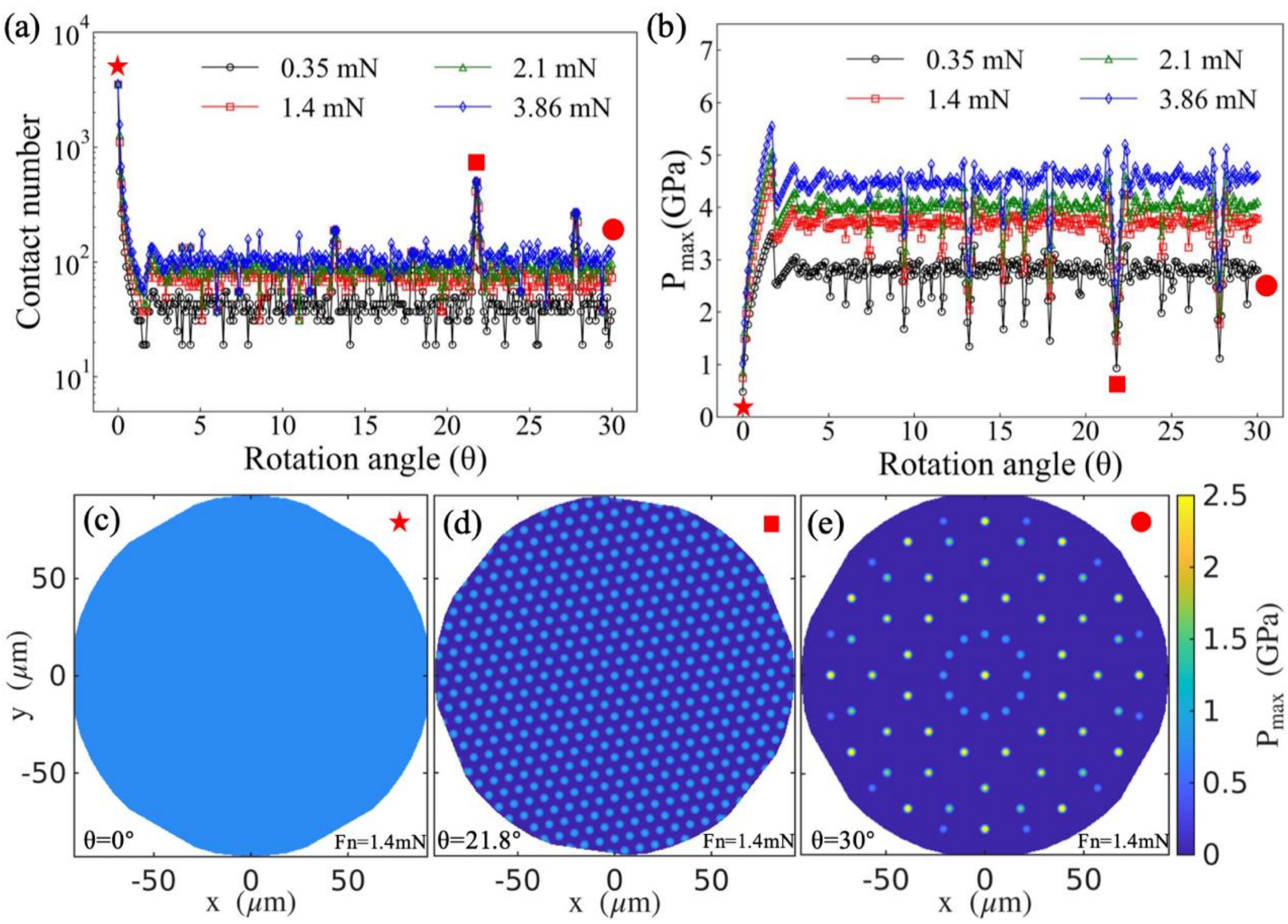

**Figure 3:** Orientation-dependent contact states of the tri-tri model.

(a) Total contact number as a function of orientation angle θ under normal loads of 0.35, 1.4, 2.1, and 3.86 mN.

(b) Maximum contact pressure ($P_{max}$) on the slider spheres as a function of orientation angle θ under the same normal loads.

(c)(d)(e) Contact configurations and pressure distributions at a normal load of 1.4 mN for representative orientations of 0°, 21.8°, and 30°, corresponding to the fully commensurate state, the incommensurate state with the largest contact number, and the typical incommensurate state, respectively.

Based on the findings above, we recognized that the robustness of structural superlubricity in this setup depends critically on balancing the number of load-bearing contacts with the incommensurability of the remaining contacts. We therefore design an interface consisting of a slider with a square lattice slider on a surface with a triangular lattice, which gives rise to a special anisotropic contact similar to the anisotropic atomic-scale structural lubricity described by Panizon *et al.*[18]

Simulations for many different orientations analogous to those described above were then performed, in which the slider orientations were varied between 0° and 30° with an angular increment of 0.1°. The cartoon model at $\theta = 0°$ is shown in **Figure 4a**. In this configuration, a centrally located commensurate contact line forms, highlighted by the green box, while the remaining parts of the slider and the substrate are not in contact, because of the mismatch between the two lattice geometries. More contact lines are expected to appear when the slider size increases to the point that the spheres at the left and right edges of the slider are positioned directly above substrate spheres. For example, at $\theta = 0°$, this condition can be realized when the slider diameter, expressed as the number of spheres multiplied by 3 μm, is approximately an integer multiple of $\sqrt{3}$.

As shown in Figure 4b and 4c, both the contact number and the maximum contact pressure ($P_{max}$) increase with increasing normal load. We focus the discussion on the range below 15° owing to the symmetries of the slider and substrate. Notably, $P_{max}$ is reduced at specific orientations, particularly 10.2° and 12.4°. Crucially, at all these orientations, including the initial configuration at $\theta = 0°$, the contact points organize into distinct straight contact lines, as illustrated in Figure 4d, 4e and **Figure S1** for $\theta = 12.4°$. In contrast, Figure 4f shows the interfacial contact pattern and pressure distribution at $\theta = 15°$, corresponding to a fully mismatched configuration without distinct contact lines. Similar to the tri-tri model, $P_{max}$ exceeds 4 GPa when the normal force increases to 3.86 mN except for the orientations close to 0°.

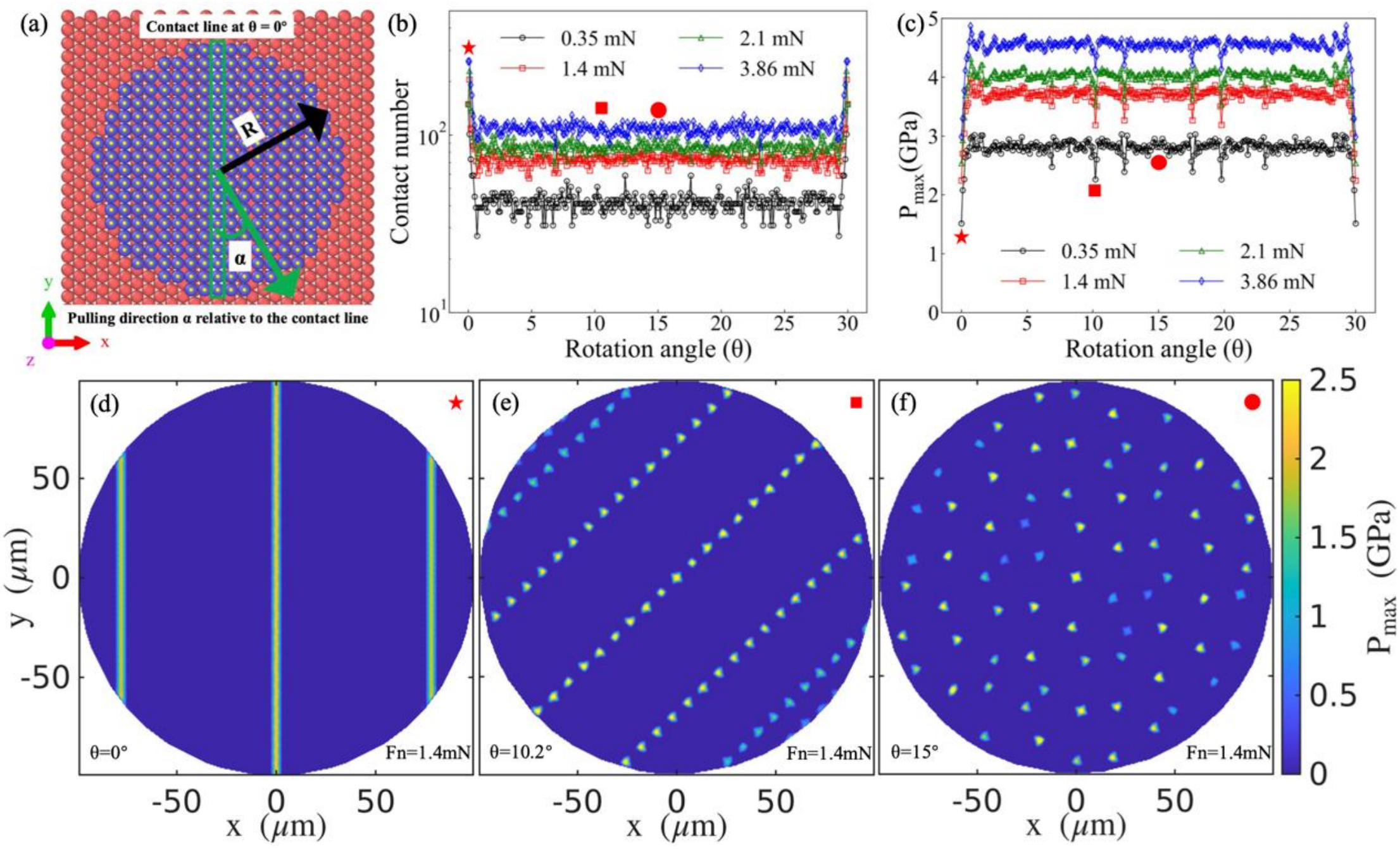

**Figure 4:** Orientation-dependent contact states of the square-triangular (square-tri) model.

(a) Top view of the cartoon square-tri model used in the orientation simulations. The dimensions of the surface, the relative orientation angle θ, and the pulling angle α are indicated. The configuration shown corresponds to θ = 0°. A centrally located commensurate contact line is highlighted by the green box.

(b) Total contact number as a function of relative orientation θ under normal loads of 0.35, 1.4, 2.1, and 3.86 mN.

(c) Maximum contact pressure ($P_{max}$) on the slider spheres as a function of orientation angle θ under the same normal loads.

(d)(e)(f) Contact configurations and pressure distributions at a normal load of 1.4 mN for representative orientations of 0°, 10.2°, and 15°, continuous contact-line state, a contact-line state with scattered contact points, and a fully mismatched configuration without distinct contact lines, respectively.

Sliding simulations were performed for the contact-line configurations at relative orientations of 0°, 10.2°, and 12.4° under normal loads of 0.35 and 1.4 mN. As there is now additional anisotropy, we examine the effect of the sliding direction angle α, defined as the angle between the pulling direction and the contact line, exploring values between 0° and 90°. Here, the corrugation is quantified by the vertical displacement of the slider along the z direction and is shown in **Figure 5a**. For θ = 0°, the corrugation remains relatively high and varies weakly with α over most angles, whereas for θ = 10.2° and 12.4°, it stays consistently low across a wide range of α. In all cases, a sharp transition occurs as α approaches 90°, where the corrugation drops abruptly to nearly zero. This indicates that sliding perpendicular to the contact lines minimizes interfacial corrugation.

This behavior can be understood from the vertical displacement of the slider on the surface as indicated in Figure 5b. The stacked lattices form continuous periodic ridges and valleys extending in the direction perpendicular to the contact lines. When the slider moves along this extension direction, that is, perpendicular to the contact lines, the sliding acts as in a one-dimensional incommensurate contact, despite a relatively large number of contacts in each contact line, similar to a one-dimensional commensurate contact. As sliding proceeds, the original contact lines gradually disengage while new ones form simultaneously. As a result, the overall height variation experienced by the slider is minimized, allowing the center of mass to glide smoothly across the surface and reducing corrugation to nearly zero.

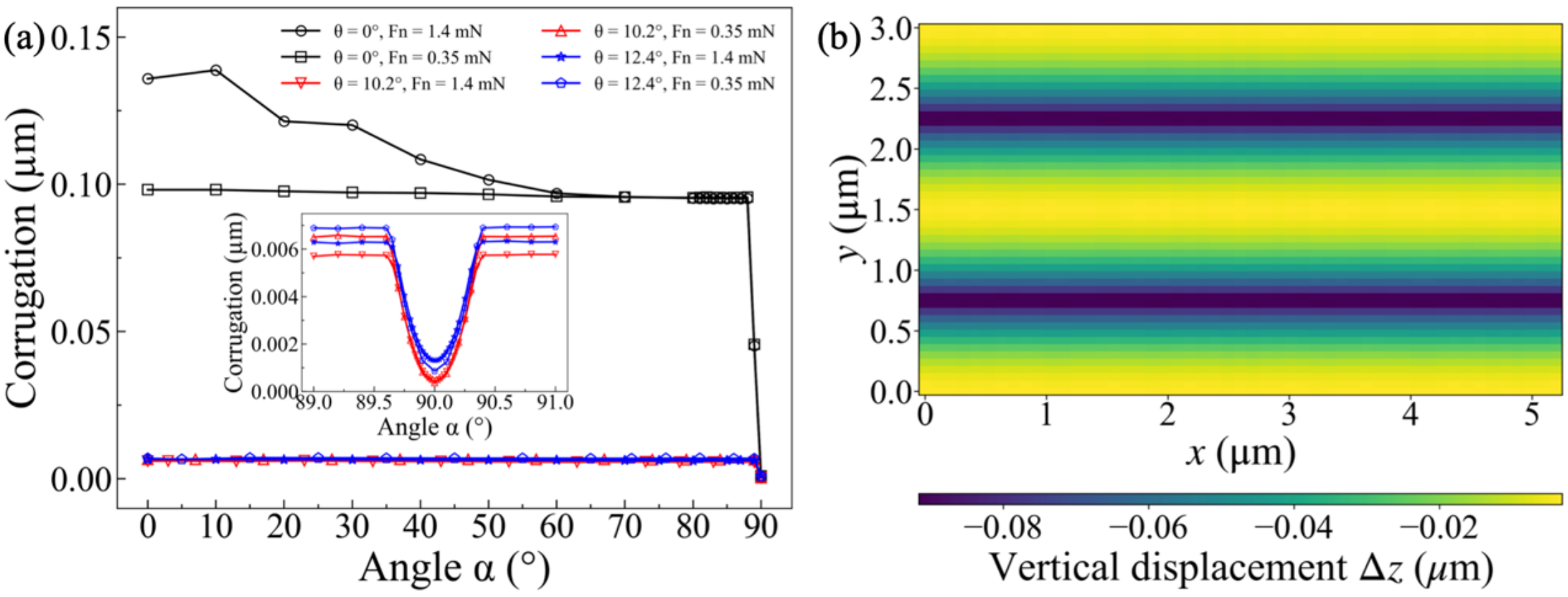

**Figure 5:** Corrugation as a function of pulling angle α and slider vertical displacement as a function of position on the surface for the square-triangular model.

(a) Corrugation amplitude (μm) of a square-lattice slider sliding on a triangular-lattice surface as a function of the pulling angle α, for three orientations, θ = 0°, 10.2°, and 12.4°, under normal loads of 0.35 and 1.4 mN. The angle α is defined as the angle between the pulling direction and the contact line, as indicated in Figure 3a. The inset shows an enlarged view of the corrugation near α = 90° for the configurations with θ = 10.2° and 12.4°.

(b) Equilibrium vertical displacement, Δz, of the square-lattice slider as a function of position on the surface, obtained by positioning the slider on a 51*51 grid spanning a 3√3*3 $\mu m^2$ rectangular repeat cell of the triangular-lattice substrate under a normal load of 1.4 mN.

We compute the friction force for the square-tri model by performing sliding simulations over the same normal-load range as that used for the tri-tri model at θ = 0°, where abundant contacts form continuous contact lines, θ = 10.2°, which exhibits lower corrugation than the θ = 12.4° configuration, and evaluate two pulling angles, α = 0° and 90°. The resulting friction forces and friction coefficients as a function of normal load for both the tri-tri and square-tri models are shown in **Figure 6a** and 6b. As discussed above for the tri-tri model, ultra-low friction is obtained and maintained only below 1.75 mN when the slider is rotated to θ = 30°, corresponding to the typical incommensurate configuration relative to the substrate. For the square-tri model at θ = 10.2°, the friction force remains slightly above zero and increases gradually with normal load when the pulling direction is parallel to the contact lines (α = 0°). In contrast, when α = 90°, i.e., when the pulling direction is perpendicular to the contact lines, the friction force remains close to zero until the normal load reaches 2.8 mN, at which point $P_{max}$ exceeds 4 GPa and the superlubricity is lost.

For the square-tri model at θ = 0°, the friction force is relatively large and increases with increasing normal load for α = 0°. However, when α = 90°, the friction force remains vanishingly small over the entire applied load range. This result shows that superlubricity can be maintained from low to relatively high loads in this specific geometric configuration, which combines contact lines with a specific pulling direction, i.e., perpendicular to the contact lines.

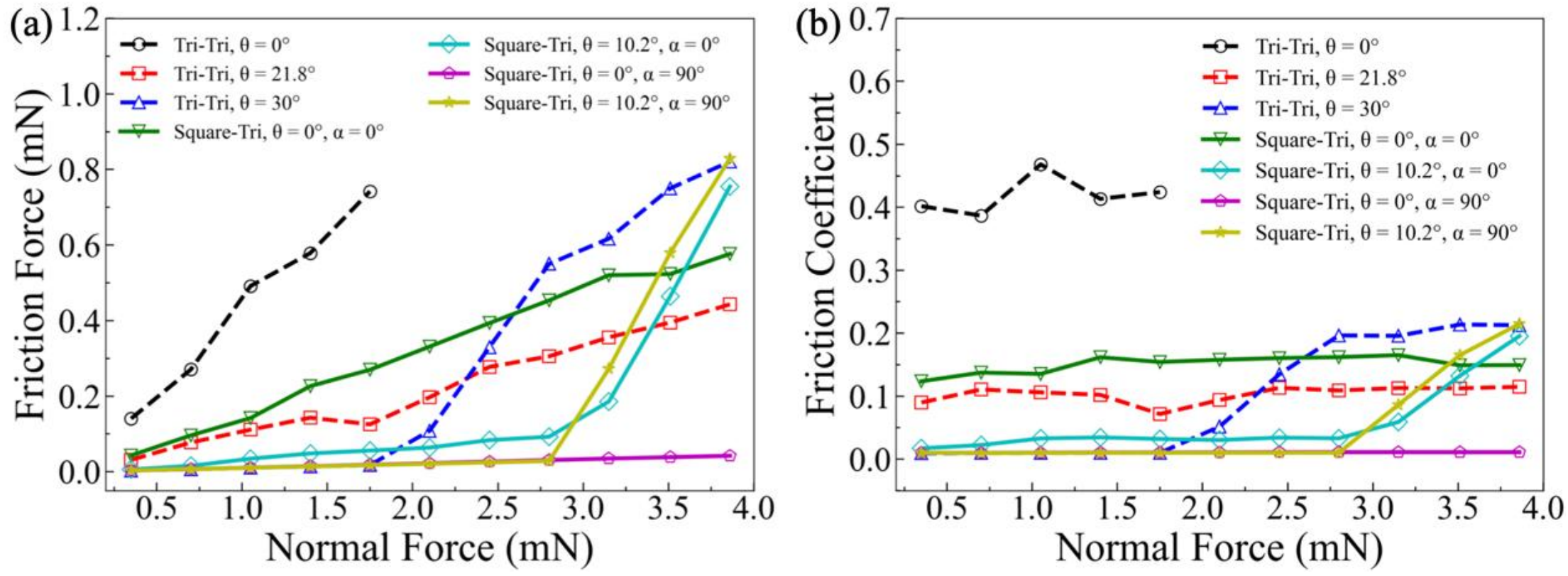

**Figure 6:** Friction force and friction coefficient as a function of normal load for the tri-tri and square-tri models.

(a) Friction force and (b) friction coefficient versus normal load from 0.35 to 3.86 mN for the tri-tri model at orientations θ = 0°, 21.8°, and 30°, and for the square-tri model at orientations θ = 0° and 10.2°. For the square-tri model, the slider was pulled at α = 0° and 90°.

In realistic manufactured surfaces, imperfections are unavoidable. To evaluate the robustness of structural superlubricity against such defects, random height variations within $\sigma_h$ = 0.002 and 0.01 μm were introduced to the spheres in the bottom surface. Their effects were examined for the tri-tri configuration at θ = 30°, where superlubricity is achieved under normal loads up to 1.75 mN, and for the square-tri configuration at θ = 0° and α = 90°, where superlubricity is maintained across the entire investigated normal-load range. The results are shown in **Figure 7**. The tri-tri configuration is highly sensitive to height variations. For spheres with a Young's modulus of 100 GPa, superlubricity breaks down under all investigated normal loads when $\sigma_h$ = 0.01 μm, and survives only below 1.05 mN when $\sigma_h$ = 0.002 μm. In contrast, the square-tri configuration exhibits substantially greater robustness. Superlubricity is maintained over the entire normal-load range for $\sigma_h$ = 0.002 μm, while for the larger variation of $\sigma_h$ = 0.01 μm, breakdown occurs only when the normal load exceeds 1.4 mN. These results demonstrate that the specific square-tri configuration can tolerate a significantly greater degree of geometric imperfection than the fully incommensurate tri-tri configuration. This enhanced tolerance arises because the square-tri configuration still retains sufficient load-bearing contacts after height variations are introduced, thereby maintaining a more uniform pressure distribution.

Since substrates in practical applications may be constructed from materials with different mechanical properties, we further investigated the effect of elasticity by performing simulations with a Young's modulus of the spheres of 50 instead of 100 GPa for the square-tri configuration with $\sigma_h$ = 0.01 μm. Under these conditions, superlubricity is recovered across the full normal-load range investigated. This enhanced robustness arises because softer spheres deform more readily under load, increasing the effective contact area and distributing interfacial pressure more uniformly, thereby protecting against local coating failure, as reported in our previous study[15]. Therefore, using softer materials is expected to further enhance tolerance to larger height variations and higher normal loads.

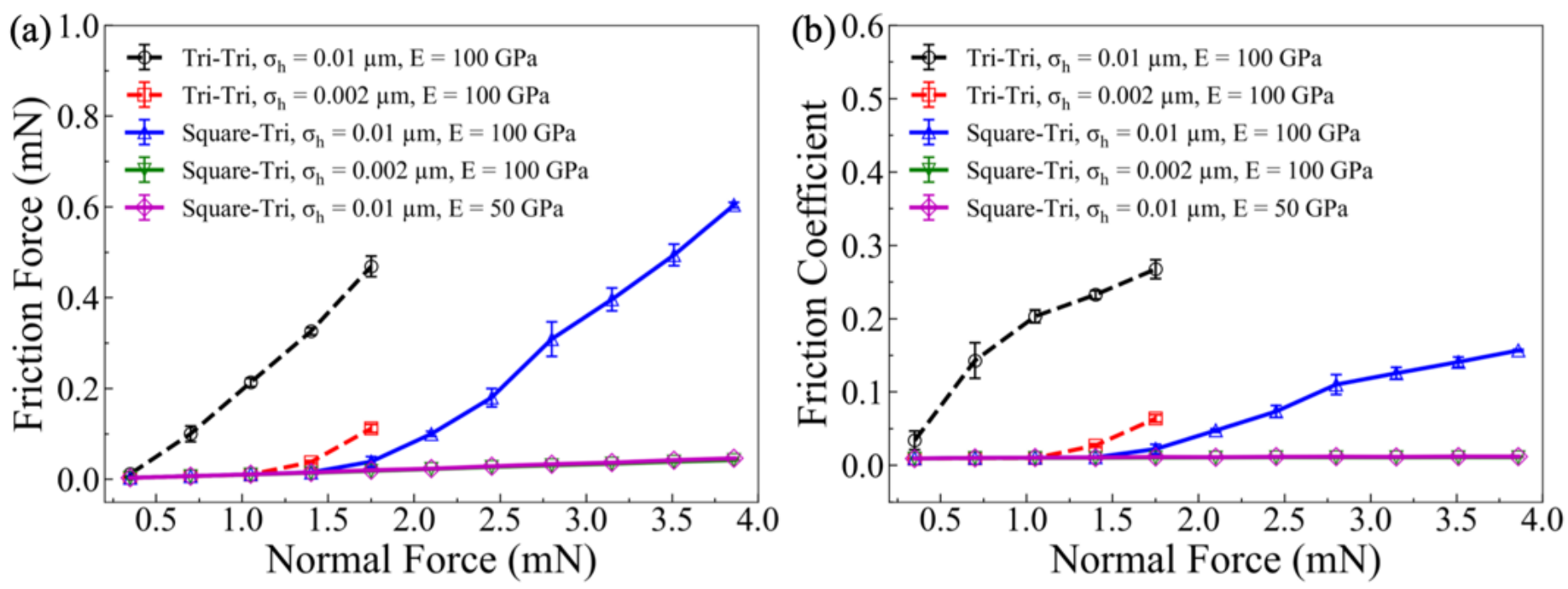

**Figure 7:** Friction force and friction coefficient as a function of normal load for the superlubric cases of the tri-tri and square-tri models under two random height variations and two Young's moduli.

(a) Friction force and (b) friction coefficient versus normal load from 0.35 to 3.86 mN for the tri-tri configuration at θ = 30° and the square-tri configuration at θ = 0°, α = 90°, under random substrate height variations ($\sigma_h$ = 0.002 and 0.01 μm) and Young's moduli of 100 and 50 GPa.

## 3. Conclusion

In the present work, we have investigated both the breakdown and robustness of structural superlubricity in patterned mesoscale interfaces. In the triangular-triangular configuration, increasing interfacial incommensurability reduces friction, but under elevated loads the relatively small number of load-bearing contacts concentrates the applied load, potentially causing the maximum local contact pressure to exceed the coating durability limit, and thus cause the coating to fail, ultimately destroying the superlubric state. This shows that incommensurability alone is insufficient to sustain superlubricity under high loads.

By contrast, the square-triangular configuration gives rise to directional linear commensurability, in which line-like commensurate contacts coexist with surrounding incommensurate regions. At specific relative orientations and pulling directions, the commensurate contact lines bear the applied load with enough contacts, while allowing sliding over low corrugation in the perpendicular direction, preventing high local pressures. This enables robust superlubricity at significantly higher loads. This same design also provides enhanced tolerance to realistic substrate imperfections, such as height variations, particularly when implemented with softer materials that more effectively redistribute interfacial pressure.

These findings reveal a previously unexplored mechanism that affects the robustness of structural superlubricity and establish a practical design principle for engineering low-friction interfaces with enhanced resistance to coating failure and wear, load-bearing capacity, and defect tolerance through mesoscale patterning and directional control.

## 4. Simulation model and methods

*Particle properties and contact model between particles*

The particle diameters were 3 μm for the substrate and slider, and 1 μm for the control layer and support. The particle density was set to 2.2 g/cm$^3$, and the particle masses were calculated from the corresponding sphere volumes. Each sphere was assigned a Young's modulus of 100 GPa and a Poisson's ratio of 0.3.

All particles were modeled as deformable elastic spheres interacting through a Hertz-Mindlin contact model with viscoelastic damping, as reported in our previous study[15]. Contact between the slider and the surface asperity particles was modeled using a Hertzian elastic contact law for normal forces combined with a modified Mindlin frictional law for tangential forces (details below). A small viscous damping term was included in the tangential contact model to account for energy dissipation, with the damping force proportional to the relative tangential sliding velocity.

To capture the breakdown of the superlubric coating at high contact pressures, we introduced a critical pressure threshold ($P_{crit}$) in the tangential model. When the local maximum Hertzian contact pressure ($P_{max}$) stays below $P_{crit}$, the interface remains in an ultra-low friction (superlubric) regime with a very low friction coefficient ($\mu_1$), but if $P_{max}$ exceeds $P_{crit}$ the coating is considered failed and the friction coefficient increases to a higher value ($\mu_2$). In our simulations, we set $P_{crit}$ = 4 GPa, $\mu_1$ = 0.01 and $\mu_2$ = 0.5 for these two regimes. More details of the modified Mindlin model are provided in our previous work[15].

Four particle types were defined to represent the substrate, slider, control layer, and support. The orientation simulations included the first three particle types, whereas the sliding simulations included all four, with the support connected to the center of mass of the control layer, and a control-layer particle for every slider particle. To ensure that the slider and control layer moved together as a single body, each pair of spheres sharing the same in-plane coordinates was connected by stiff directional springs along the x, y, and z directions, with stiffness coefficients of $10^3$, $10^3$, and $10^6$ N/m, respectively. The control layer was then constrained as a rigid body. A normal load was applied to the control layer and transmitted to the underlying slider via the spring. The center of mass of the control layer was coupled to a support particle that moved at a constant velocity.

The substrate spheres were tethered to their initial positions by harmonic springs in the lateral x and y directions, with stiffness coefficients $K_x$ and $K_y$ of $10^3$ N/m, while remaining fixed in the normal z direction. To prevent tilting of the slider, the control layer was divided into four regions, and opposing regions were coupled via harmonic springs acting in the normal (z) direction. The left-right and upper-lower halves were connected through center-of-mass spring coupling ($K_1 = 10^6$ N/m), suppressing rotation about the x and y axes while allowing in-plane translation of the slider.

*Geometric configuration*

The lattice spacing of the triangular lattice was defined as $a = 3\sqrt{3}$ µm, whereas that of the square lattice was $b = 3$ µm, consistent with a particle diameter of 3 µm. The radii of the circular triangular-lattice and square-lattice sliders were 93.5 and 100 µm, respectively, so that both contained the same number of spheres in the slider, specifically 3505.

The rectangular-shaped substrate used in all simulations had dimensions of 364 µm × 270 µm. The longer side was aligned with the sliding direction to ensure a sufficiently long trajectory for stable dynamic-friction measurements. In the sliding simulations, the support was positioned 120 µm along the x direction from the slider and connected to the center of mass of the control layer via a spring ($K_0 = 50$ N/m).

*Simulation protocol*

All simulations were performed in the Large-scale Atomic/Molecular Massively Parallel Simulator (LAMMPS)[19] using the Granular package with custom modifications developed by us[15].

The system was integrated using an NVE scheme with a Langevin thermostat at 0 K to dissipate residual kinetic energy and a viscous damping with a damping time of 1 µs. All simulations used a time step of $10^{-5}$ µs.

For the orientation study, the system was equilibrated for 10 μs at each orientation of the slider under the prescribed normal load. The contact number and maximum contact pressure ($P_{max}$) were extracted from the final equilibrated configuration.

For the sliding simulations at selected orientations, three sequential stages were performed: equilibrium, loading, and sliding. The system was first equilibrated for 10 μs under an NVE ensemble with Langevin damping as described above to remove residual kinetic energy. The slider was then coupled to an external support via a harmonic spring and further relaxed under the applied normal force for an additional 10 μs. Sliding was initiated by moving the support at a constant velocity of 0.1 m/s for 300 μs. The friction force was obtained from the spring force during steady state sliding by averaging over the interval from 100 to 300 μs.

**Acknowledgements**

This work was funded by the HORIZON-EIC-2021-PATHFINDEROPEN-01 through the project “SSLiP: Scaling-up Superlubricity into persistence” (Project No. 101046693). The simulations were performed using resources provided by the IDUN HPC cluster at the Norwegian University of Science and Technology (NTNU) and Sigma2-the National Infrastructure for High-Performance Computing and Data Storage in Norway (Project No. NN10020K). Views and opinions expressed are however those of the author(s) only and do not necessarily reflect those of the European Union or EIC. Neither the European Union nor the granting authority can be held responsible for them.

**Data Availability statement**

The data supporting the findings of this study are available in the article and its supplementary materials. All data are available from the corresponding author upon request.

**Supporting Information**

Supporting Information is available from the Wiley Online Library.

**Author contributions**

A. S. d. W., G. L. W. C., and V. H. H. conceived the idea. V. H. H., G. L., and M. M. G. developed the model. G. L. and V. H. H. conducted the simulations and analysed the results. G. L. prepared the original draft. A. S. d. W., G. L. W. C., and B. H. supervised the research. All authors discussed the results and commented on the manuscript.

**Competing interests**

The authors declare no competing interests.

Patterned surfaces coated with a superlubric layer can reduce friction through incommensurate patterns. However, this leaves sparse contacts to carry the load, causing locally high pressures and coating failure. A square-on-triangular pattern creates directional commensurability that allows for abundant contacts along the commensurate direction and low local loads while also maintaining low friction for sliding in the incommensurate direction.

**Directional commensurability stabilizes structural superlubricity in patterned mesoscale interfaces**

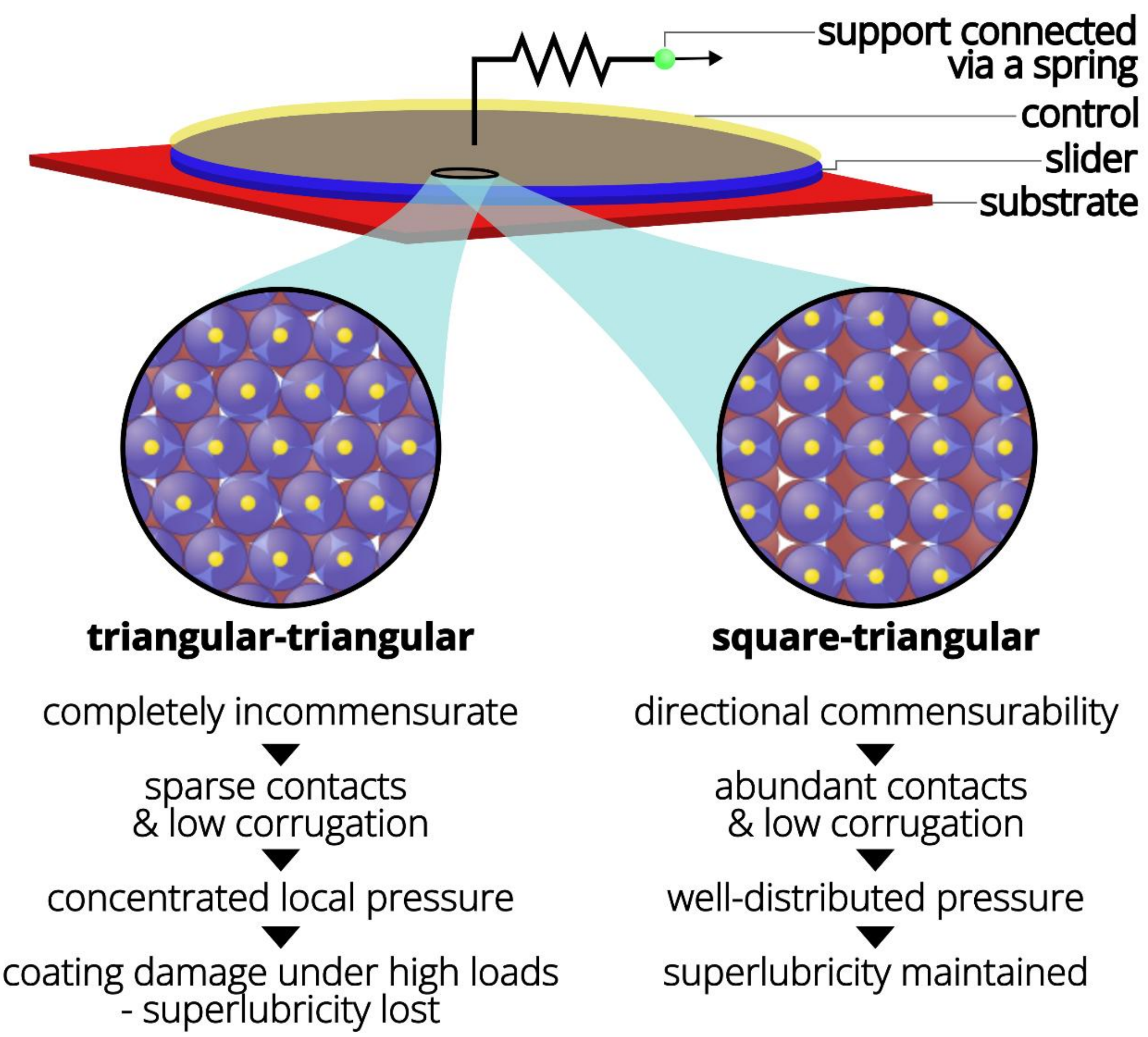